\documentclass[10pt,a4paper,twocolumn,showpacs,showkeys,lengthcheck,nobibnotes,prd,balancelastpage]{revtex4}
\usepackage{amsfonts}
\usepackage{amsmath}
\usepackage{amssymb}
\usepackage{graphicx}

\begin{document}

\title[ ]{\textbf{Quantized self-intervening detector networks}}
\author{George Jaroszkiewicz}
\affiliation{School of Mathematical Sciences, University of Nottingham, Nottingham, UK}

\begin{abstract}
A range of quantum optics experiments is discussed in which the apparatus
can be modified by detector outcomes during the course of any run. Starting
with a single beamsplitter network, we work our way through a series of more
complex scenarios, culminating with a proposed self-intervening experiment
which could provide evidence for the existence of the Heisenberg cut, the
supposed boundary between classical and quantum physics.
\end{abstract}

\date{\today}
\maketitle

\section{Introduction}

Many quantum experiments involve time-independent apparatus. By this we mean
that for each run of such an experiment, the apparatus which prepares the
initial state, shields it from the environment during that run and detects
the outcome state is classically determined and fixed. Such experiments will
be called Type 1. In quantum optics, double-slit, Mach-Zehnder, and quantum
eraser experiments \cite{WALBORN-2002} are of this type. So too are high
energy particle scattering experiments such as those conducted at the LHC in
CERN. Type 1 experiments are important because they allow the focus of
attention to be entirely on the dynamical evolution of states of SUOs
(systems under observation), normally regarded as the prime objective of SQM
(standard quantum mechanics). SQM generally describes Type 1 experiments via
time-independent Hamiltonians.

A second class of experiment, referred to here as Type 2, involves some
time-dependence in the apparatus, such that any changes in the apparatus
during a run are controlled externally, either by the experimentalist or by
environmental factors. Spin-echo magnetic resonance experiments are of this
type, because the experimentalist arranges for certain magnetic fields to be
rotated precisely whilst additionally, the environment introduces random
external influences related to local temperature. An example of random
changes controlled by the experimentalist are delayed-choice experiments
such as that of Jacques et al \cite{JACQUES+AL-2006}, where carefully
arranged random changes are made during each run. SQM typically describes
such experiments via time-dependent Hamiltonians.

Type 2 experiments are more interesting than Type 1 because they have the
potential to reveal more information about the dynamics of SUOs than Type 1.
Schwinger's source theory shows that in principle, Type 2 experiments allow
for the extraction of all possible information about quantum systems \cite%
{SCHWINGER-1969}. Type 1 and Type 2 experiments may be collectively labelled
as \emph{exophysical}, because all classical apparatus interventions are
external in origin. In such experiments, the apparatus is classically well
defined at each instant of time during each run, even in those situations
where it changes randomly. Therefore, a classical block universe \cite%
{PRICE:1997} account of apparatus during each run of a Type 1 or 2
experiment is possible.

In this paper we explore a third type of quantum experiment, which we label
Type 3, or \emph{endophysical.} In such experiments, the apparatus is
modified internally by the quantum dynamics of the SUO, rather than
externally by the observer or the environment. An interesting question which
we shall address towards the end of this paper is whether Type 3 experiments
can always be given a classical block universe account or whether something
analogous to superpositions of different apparatus has to be envisaged (not
to be confused with superpositions of states of SUOs).

This question is related to the rules of quantum information extraction as
they are currently known in SQM. These rules state that quantum interference
can occur in the absence of classical which-path information, the most
well-known example of this being the double-slit experiment. The question
here is what precisely does a lack of which-path information mean: \emph{if}
such as thing as a photon passed through one of the slits, would it leave
any trace in principle? Even if it did, it might be believed that any such
interaction a photon had with atoms at either slit would be on the quantum
level, far below the scales of classical mechanical detection, and so the
observer of the interference pattern would simply be unaware of such
interaction. This seems wrong to us on two counts: first, there is now
sufficient evidence against the notion that photons are particles in the
conventional sense \cite{PAUL:2004} and second, one observer being unaware
of actual which-path information held by another observer could not by
itself induce interference patterns. There has to be something deeper than
that in the origin of quantum interference.

The neutron interference experiment discussed by Greenberger and YaSin \cite%
{GREENBERGER+YASIN-1989} explores this question by moving towards larger
scales of interaction between SUO and apparatus. In their experiment, the
movement of mirrors involved in their quantum erasure scheme involves a
macroscopic numbers of atoms and molecules \cite{BECKER-1998}. In this case,
the dynamical effects of the impact of a particle on a mirror is reversed by
a second impact. What is amazing is the idea that all possible traces of the
first impact could be completely erased, even though there could (in
principle) be time for information from the first impact to be dissipated
into the environment, thereby rendering the process irreversible.

We take this line of thinking one step further. One of the experiments we
propose and discuss here appears to involve the superposition of states of
different beamsplitters, which are macroscopic pieces of apparatus. At least
one of these beamsplitters has to be triggered if any interference effects
are observed, but that observation cannot occur if the information as to
which beamsplitter is involved can be extracted by the observer. If such an
experiment were carried out and quantum interference observed, then the
implications would be that quantum principles apply to apparatus as well as
states of SUOs, thereby demonstrating that the laws of quantum information
are truly universal.

We focus exclusively on linear quantum evolution, i.e., one conforming with
the principles of SQM as discussed for example in \cite{PERES:1993}, rather
than appeal to any form of non-linear quantum mechanics to generate
self-intervention effects. We explore a number of Type 3 thought experiments
involving photons, which act as either quantum or classical objects at
various times. As quantum objects they pass through beam-splitters and
suffer random outcomes as a result. As classical objects they are used to
trigger the switching on or off of macroscopic apparatus, a switching which
determines the subsequent quantum evolution of other photons. We shall not
discuss the nature of photons per se, except to say that they are referred
to as particles for convenience only: our ideology and formalism treats them
as signals in elementary signal detectors (ESDs) \cite{J2007C}. Everything
is idealized here, it being assumed that all detectors operate with one
hundred percent efficiency and that photon polarizations and wavelengths can
be adjusted wherever necessary to make the scenarios discussed here
physically realizable. The experiments we discuss are not necessarily based
on photons: other particles such as electrons could be used in principle. We
use the Schr\"{o}dinger picture throughout, using a Hilbert space quantum
register of sufficiently many qubits to model all information exchange
requirements. In our notation, $bc$ denotes a two photon signal state,
equivalent to $|b\rangle \otimes |c\rangle $ in standard notation and to $%
\mathbb{A}_{b}^{+}\mathbb{A}_{c}^{+}|0)$ in \cite{J2007C}, where $|0)$ is
the void or $``$no-signal$"$ state of the apparatus and $\mathbb{A}_{b}^{+}$
is a signal operator creating a positive signal state at ESD $b$. Capital
letters such as $E_{1}$ represent complex outcome probability amplitudes.

\section{Experiment 1: basic self-intervention}

To illustrate the sort of experiment we are interested in, we start with the
basic experiment shown schematically in Figure 1. A correlated,
non-entangled two photon state $\Psi _{0}\equiv bc$ is created by source $A$%
. Such states can be created by parametric down conversion and suitable
filtering. Photon $c$ is subsequently passed through beamsplitter $D$ and
emerges in state $d_{1}$ or $d_{2}$ with amplitudes $D_{1}$ and $D_{2}$
respectively, such that $|D_{1}|^{2}+|D_{2}|^{2}=1$. If ESD $d_{1}$ is
triggered rather than ESD $d_{2}$, then a macroscopic mechanism triggers
beamsplitter $E$ to be activated. In all diagrams, squares denote apparatus
modules such as sources of photon pairs and beamsplitters, circles denote
ESDs and single lines denote optical pathways. Dotted double lines denote
signal detection followed by classical switching on of apparatus.

Photon $b$ meanwhile is sent over a sufficiently long optical path to ensure
that photon $c$ has been detected in one of the ESDs $d_{1}$ or $d_{2}$.
Only then does $b$ enter that part of the apparatus which has been prepared
by the outcome of beamsplitter $D$. If the outcome was $d_{1}$, then $b$
enters beamsplitter $E$ and emerges in state $e_{1}$ or $e_{2}$ with
amplitude $E_{1}$ and $E_{2}$ respectively, such that $%
|E_{1}|^{2}+|E_{2}|^{2}=1$. On the other hand, if the outcome at $D$ was $%
d_{2}$, then $E$ is not switched on, so that $b$ is unaffected and gets
registered as an unaltered photon $b$.

\begin{figure}[t]
\centerline{\includegraphics[width=2in]{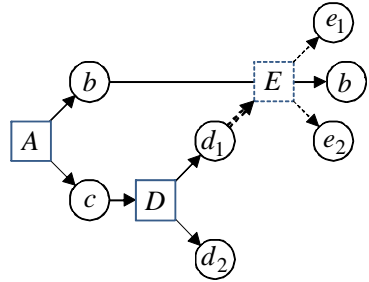}}
\caption{If detected, photon $d_{1}$ triggers the switching on of
beamsplitter $E$.}
\label{fig:1}
\end{figure}

The labstate \cite{J2007C} $\Psi _{1}$ just before any photons are detected
is given by%
\begin{equation}
\Psi _{1}=D_{1}(E_{1}e_{1}+E_{2}e_{2})d_{1}+D_{2}bd_{2}.  \label{111}
\end{equation}%
In (\ref{111}), photon $d_{1}$ is included in the state explicitly. This is
because although it is necessarily absorbed during the switching on of
beamsplitter $E$, the observer can determine the fact that that switching
has occurred, and this is equivalent to the detection of a photon by an ESD.
In our formalism, ESDs are any processes which result in classical signal
information being extracted from a quantum state. As stated above, we think
of photons not as particles but as quanta of information.

From (\ref{111}) we can immediately read off the three possible non-zero
outcome probabilities:%
\begin{eqnarray}
P(e_{1}\&d_{1}) &=&|D_{1}|^{2}|E_{1}|^{2},\ \
P(e_{2}\&d_{1})=|D_{1}|^{2}|E_{2}|^{2},  \notag \\
P(b\&d_{2}) &=&|D_{2}|^{2},
\end{eqnarray}%
which sum to unity as required.

\section{Experiment 2: double self-intervention}

The next variant experiment is shown in Figure 2. Now photons $b$ and $c$
pass through beamsplitters $E$ and $D$ respectively. If detected, outcome $%
d_{1}$ of $D$ switches on beamsplitter $F$, whereas if detected, outcome $%
d_{2}$ of $D$ switches on beamsplitter $G$.

The dynamics is calculated as follows. The initial labstate is $\Psi _{0}=bc$%
. Subsequently, we have $b\rightarrow (E_{1}e_{1}+E_{2}e_{2})$ and $%
c\rightarrow (D_{1}d_{1}+D_{2}d_{2})$, with $%
|E_{1}|^{2}+|E_{2}|^{2}=|D_{1}|^{2}+|D_{2}|^{2}=1$. Hence at stage one, the
labstate is $\Psi _{1}=(E_{1}e_{1}+E_{2}e_{2})(D_{1}d_{1}+D_{2}d_{2})$.

\begin{figure}[t]
\centerline{\includegraphics[width=3in]{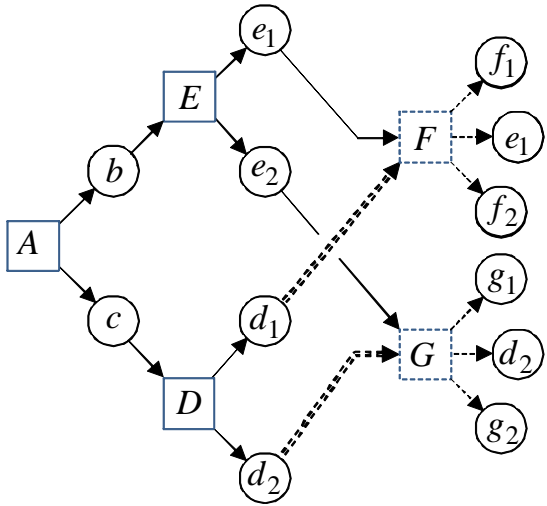}}
\caption{In this scheme, no quantum interference occurs.}
\label{fig:2}
\end{figure}

For the next stage, we refer to Figure $2$ to write down the following
substage evolution rules:%
\begin{eqnarray}
e_{1}d_{1} &\rightarrow &(F_{1}f_{1}+F_{2}f_{2})d_{1},\ \ \
e_{1}d_{2}\rightarrow e_{1}d_{2},  \notag \\
e_{2}d_{2} &\rightarrow &(G_{1}g_{1}+G_{2}g_{2})d_{2},\ \ \
e_{2}d_{1}\rightarrow e_{2}d_{1},
\end{eqnarray}%
where $|F_{1}|^{2}+|F_{2}|^{2}=|G_{1}|^{2}+|G_{2}|^{2}=1$. This gives%
\begin{eqnarray}
\Psi _{2} &=&E_{1}D_{1}(F_{1}f_{1}+F_{2}f_{2})d_{1}+E_{1}D_{2}e_{1}d_{1}
\notag \\
&&+E_{2}D_{1}e_{2}d_{1}+E_{2}D_{2}(G_{1}g_{1}+G_{2}g_{2})d_{2}.
\end{eqnarray}%
From this we immediately read off six non-zero correlation probabilities,
such as $P(f_{1}\&d_{1})=|E_{1}|^{2}|D_{1}|^{2}|F_{1}|^{2}$, and so on. None
of these probabilities demonstrates any quantum interference, because
complete which-path information is available in each case.

\section{Experiment 3: interfering single self-intervention}

The third scenario is shown in Figure 3. In this case, the initial photon
pair is passed through a pair of beamsplitters exactly as in Experiment 2.
The difference lies in the next stage. Photons $e_{1}$ and $d_{2}$ are sent
off over sufficiently long optical paths so as to allow interference between
photons $e_{2}$ and $d_{1}$ in beamsplitter $F$. Note that this interference
essentially involves waves from different source photons $b$ and $c$, so
phase and wavelength matching would have to be arranged. Provided this is
the case, then the dynamics is given by%
\begin{align}
e_{1}d_{1}& \rightarrow e_{1}(F_{1}f_{1}+F_{2}f_{2}),\ \ \
e_{1}d_{2}\rightarrow e_{1}d_{2}  \notag \\
e_{2}d_{2}& \rightarrow (F_{3}f_{1}+F_{4}f_{2})d_{2},\ \ \
e_{2}d_{1}\rightarrow e_{2}d_{1}
\end{align}%
where the coefficients $\{F_{i}\}$ satisfy the relations%
\begin{equation}
|F_{1}|^{2}+|F_{2}|^{2}=|F_{3}|^{2}+|F_{4}|^{2}=1,\ \ \ F_{1}F_{3}^{\ast
}+F_{2}F_{4}^{\ast }=0,
\end{equation}%
these being the most general conditions required to ensure unitarity. The
labstate at this stage is therefore%
\begin{align}
\Psi _{2}=& E_{1}D_{1}e_{1}(F_{1}f_{1}+F_{2}f_{2})+E_{1}D_{2}e_{1}d_{2}+
\notag \\
& E_{2}D_{1}e_{2}d_{1}+E_{2}D_{2}(F_{3}f_{1}+F_{4}f_{2})d_{2}.
\end{align}%
If the outcome from $F$ is $f_{1}$, then beamsplitter $G$ is switched on,
otherwise it remains off.

\begin{figure}[t]
\centerline{\includegraphics[width=3.5in]{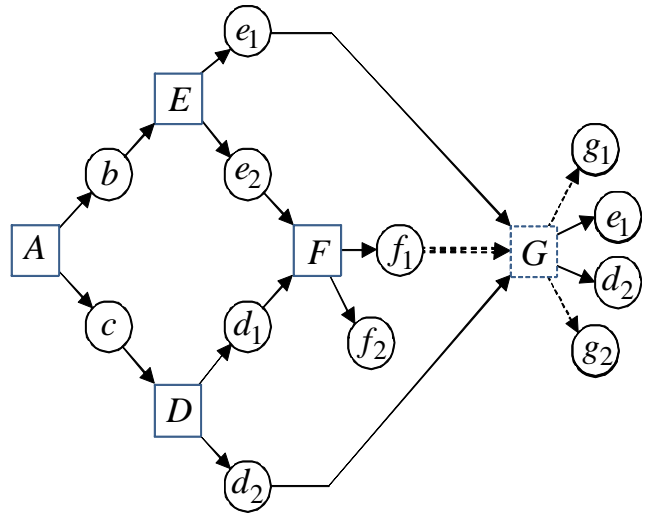}}
\caption{Interference at beamsplitter $F$ can trigger beamsplitter $G$.}
\label{fig:3}
\end{figure}

For the final stage, photons $e_{1}$ and $d_{2}$ are brought together so as
to interact with the result of the decision made at $F$. If $f_{1}$ had been
triggered, then $e_{1}$ and $d_{2}$ interact in beamsplitter $G$, otherwise
they evolve normally and do not interact. The dynamics is given by%
\begin{align}
e_{1}f_{1}& \rightarrow (G_{1}g_{1}+G_{2}g_{2})f_{1},\ \ \
e_{1}f_{2}\rightarrow e_{1}f_{2},\ \ \ e_{1}d_{2}\rightarrow e_{1}d_{2}
\notag \\
f_{1}d_{2}& \rightarrow f_{1}(G_{3}g_{1}+G_{4}g_{2}),\ \ \
e_{2}d_{1}\rightarrow e_{2}d_{1},\ \ \ f_{2}d_{2}\rightarrow f_{2}d_{2},
\end{align}%
where the $\{G_{i}\}$ coefficients satisfy the generalized beamsplitter
relations%
\begin{equation}
|G_{1}|^{2}+|G_{2}|^{2}=|G_{3}|^{2}+|G_{4}|^{2}=1,\ \ \ G_{1}G_{3}^{\ast
}+G_{2}G_{4}^{\ast }=0.
\end{equation}

The final labstate is given by%
\begin{align}
\Psi _{3}=&
(E_{1}D_{1}F_{1}G_{1}+E_{2}D_{2}F_{3}G_{3})f_{1}g_{1}+E_{1}D_{1}F_{2}e_{1}f_{2}
\notag \\
& +E_{1}D_{2}e_{1}d_{2}+(E_{1}D_{1}F_{1}G_{2}+E_{2}D_{2}F_{3}G_{4})f_{1}g_{2}
\notag \\
& +E_{2}D_{2}F_{4}f_{2}d_{2}+E_{2}D_{1}e_{2}d_{1},
\end{align}%
from which we read off the non-zero probabilities%
\begin{align}
P(f_{1}\&g_{1})& =|E_{1}D_{1}F_{1}G_{1}+E_{2}D_{2}F_{3}G_{3}|^{2},  \notag \\
P(f_{1}\&g_{2})& =|E_{1}D_{1}F_{1}G_{2}+E_{2}D_{2}F_{3}G_{4}|^{2},  \notag \\
P(e_{1}\&f_{2})& =|E_{1}D_{1}F_{2}|^{2},\ \ \
P(e_{1}\&d_{2})=|E_{1}D_{2}|^{2},  \notag \\
P(d_{2}\&f_{2})& =|E_{2}D_{2}F_{4}|^{2},\ \ \
P(e_{2}\&d_{1})=|E_{2}D_{1}|^{2}.
\end{align}%
These probabilities sum to unity as required.

In this variant experiment, two of the outcome probabilities, $%
P(f_{1}\&g_{1})$ and $P(f_{1}\&g_{2})$ show interference. This interference
has essential contributions from beamsplitters $F$ and $G$ in a manner which
seems impossible to explain in terms of photons as classical particles.

In all experiments where quantum interference takes place, there inevitably
has to be some which-path uncertainty somewhere. This does not occur in
variant experiments $1$ or $2$ but does occur in variant $3$.

The results can be simplified somewhat by assuming each beamsplitter is
symmetric, i.e., $|E_{1}|=|E_{2}|=1/\sqrt{2}$. etc. There is a well-known
change of phase due to reflection at a beamsplitter, relative to the
transmitted beam \cite{ZEILINGER-1981}. If we take $%
D_{1}=E_{1}=F_{1}=F_{4}=G_{1}=G_{4}=1/\sqrt{2}$, $%
D_{2}=E_{2}=F_{2}=F_{3}=G_{2}=G_{3}=i/\sqrt{2}$, we find
\begin{align}
P(f_{1}\&g_{1})& =\frac{1}{4},\ \ \ P(e_{1}\&f_{2})=\frac{1}{8},\ \ \
P(e_{1}\&d_{2})=\frac{1}{4},  \notag \\
P(f_{1}\&g_{2})& =0,\ \ \ P(d_{2}\&f_{2})=\frac{1}{8},\ \ \ P(e_{2}\&d_{1})=%
\frac{1}{4}.
\end{align}%
In an actual experiment, we expect that pathlengths would need to be tuned
carefully in order to obtain these effects. Rotating beamsplitter $F$ so as
to interchange the roles of reflection and transmission at $F$ should then
interchange the results for $P(f_{1}\&g_{1})$ and $P(f_{1}\&g_{2})$,
confirming that constructive and destructive interference is taking place.
Similar remarks apply to beamsplitter $G$.

Proposed Experiment 3 should be viable with current technology. If
interference effects were detected as predicted, then that would demonstrate
not only that classical information extraction (at $f_{1}$) need not destroy
interference taking place after that extraction, but also that such
classical intervention can play an essential role in the optical paths
involved.

Although it involves the possibility of self-intervening apparatus change,
experiment 3 does not involve any erasure of such a change. To investigate
this, we need to go further. To this end, we first return to the basic
double slit experiment and investigate what happens when we try to detect a
photon at any of the slits.

\section{Experiment 4: double-slit which-way detection}

The double-slit experiment with no which-path detection is shown in Figure 4.

\begin{figure}[t]
\centerline{\includegraphics[width=2in]{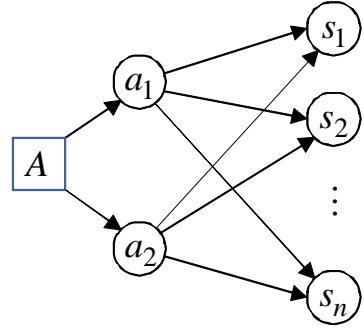}}
\caption{The double-slit experiment.}
\label{fig:4}
\end{figure}

The initial labstate is $\Psi _{0}=A_{1}a_{1}+A_{2}a_{2}$, where $%
|A_{1}|^{2}+|A_{2}|^{2}=1$, and the detection screen consists of $n$ ESDs $%
\{s_{i}:i=1,2,\ldots ,n\}$, with $n\geqslant 2$. The dynamical rules for
no-which-path detection are%
\begin{equation}
a_{1}\rightarrow \sum_{i=1}^{n}S_{i}s_{i},\ \ \ a_{2}\rightarrow
\sum_{i=1}^{n}T_{i}s_{i},  \label{222}
\end{equation}%
where%
\begin{equation}
\sum_{i=1}^{n}|S_{i}|^{2}=\sum_{i=1}^{n}|T_{i}|^{2}=1,\ \ \
\sum_{i=1}^{n}S_{i}T_{i}^{\ast }=0.
\end{equation}%
Hence the final labstate is%
\begin{equation}
\Psi _{1}=\sum_{i=1}^{n}\{A_{1}S_{i}+A_{2}T_{i}\}s_{i},
\end{equation}%
from which we read off the detection probabilities%
\begin{equation}
P(s_{i})=|A_{1}S_{i}+A_{2}T_{i}|^{2},\ \ \ i=1,2,\ldots ,n.
\end{equation}%
These demonstrate quantum interference and sum to unity as required.

Now suppose that we allow for the possibility of detecting from which slit a
photon came as it lands on the detecting screen. We introduce two new ESDs,
labelled $u$ and $v$, which give information about $a_{1}$ and $a_{2}$
respectively. The experimental architecture is now given by Figure 5.

\begin{figure}[h!]
\centerline{\includegraphics[width=2in]{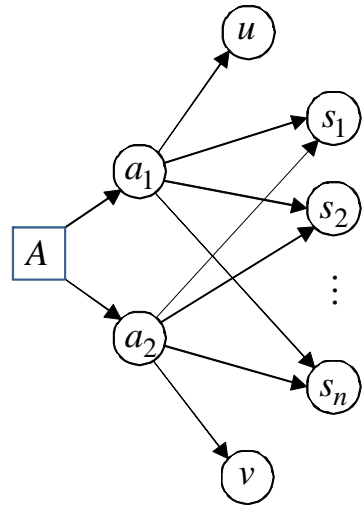}}
\caption{Double-slit which-path detection.}
\label{fig:5}
\end{figure}

Assuming perfectly efficient detection at $u$ and $v$, the dynamics is now
given by%
\begin{equation}
a_{1}\rightarrow \sum_{i=1}^{n}S_{i}s_{i}u,\ \ \ a_{2}\rightarrow
\sum_{i=1}^{n}T_{i}s_{i}v,  \label{333}
\end{equation}%
so the final labstate is now%
\begin{equation}
\Psi _{2}=\sum_{i=1}^{n}\{A_{1}S_{i}s_{i}u+A_{2}T_{i}s_{i}v\}.
\end{equation}%
This time, we have the probabilities%
\begin{equation}
P(u\&s_{i})=|A_{1}S_{1}|^{2},\ \ \ P(v\&s_{i})=|A_{2}T_{i}|^{2},
\end{equation}%
which sum to unity. Moreover, the total probability $P(s_{i})$ is just the
sum $P(u\&s_{i})+P(v\&s_{i})$, which is the classical expectation showing no
interference.

We can imagine performing a variant of this experiment where detection at
each slit is not perfect, i.e., we consider a smooth transition from the
complete which-path scheme of Figure $5$ to the complete no-which-path
scheme of Figure $4$. We replace the dynamical schemes (\ref{222}) or (\ref%
{333}) by%
\begin{eqnarray}
a_{1} &\rightarrow &\sum_{i=1}^{n}S_{i}\{\cos (\theta _{1})+\sin (\theta
_{1})u\}s_{i},  \notag \\
a_{2} &\rightarrow &\sum_{i=1}^{n}T_{i}\{\cos (\theta _{2})+\sin (\theta
_{2})v\}s_{i},
\end{eqnarray}%
where $\theta _{1}$ and $\theta _{2}$ are real. The case $\theta _{1}=\theta
_{2}=0$ corresponds to complete no-which-path information, Figure $4,$
whilst $\theta _{1}=\theta _{2}=\pi /2$ corresponds to complete which-path
information, Figure $5$. The final labstate is now given by%
\begin{eqnarray}
\Psi _{1} &=&A_{1}\sin (\theta _{1})\sum_{i=1}^{n}S_{i}us_{i}+A_{2}\sin
(\theta _{2})\sum_{i=1}^{n}T_{i}vs_{i}  \notag \\
&&+\sum_{i=1}^{n}\{A_{1}S_{i}\cos (\theta _{1})+A_{2}T_{i}\cos (\theta
_{2})\}s_{i}
\end{eqnarray}%
The respective probability distributions are readily read off and the total
detection probability $P(s_{i})$ at ESD $s_{i}$ turns out to be%
\begin{eqnarray}
P(s_{i}) &=&\{A_{1}A_{2}^{\ast }S_{i}T_{i}^{\ast }+A_{1}^{\ast
}A_{2}S_{i}^{\ast }T_{i}\}\cos (\theta _{1})\cos (\theta _{2})  \notag \\
&&+|A_{1}S_{i}|^{2}+|A_{2}T_{i}|^{2}.
\end{eqnarray}%
This shows how the quantum interference term disappears the more which-path
information becomes certain, i.e., in the limits when $\theta
_{1}\rightarrow \pi /2$ or $\theta _{2}\rightarrow \pi /2$. A particular
feature of the double-slit experiment is that detection at just one slit
alone destroys the interference term. Another important feature is that
elimination of which-way information is pre-determined by the observer
choosing not to have any detection equipment at both slits, corresponding to
$\theta _{1}=\theta _{2}=0$, rather than any quantum erasure process of the
type discussed in recent experiments \cite{WALBORN-2002}. What is surprising
is that this choice is quite sufficient to produce interference on the
detecting screen even in the case where the particles are bound states such
as neutrons or molecules and cannot be regarded as elementary in any true
sense.

Before the advent of quantum mechanics, any analysis suggesting that lack of
which-path information about particle trajectories led to interference would
have been regarded as fanciful. Yet double-slit interference is an
experimental fact. It is reasonable, therefore, to investigate the
possibility that such interference could happen on even larger macroscopic
scales, provided true which-way information was absent. To this end we
propose the experiment discussed next.

\section{Experiment 5: interfering beamsplitters}

Consider the detector network shown in Figure 6, a modification of the
double self-intervention network shown in Figure 2. The difference is that
the triggered beamsplitters $F$ and $G$ now feed onto the same pair of ESDs
labelled $f_{1}$ and $f_{2}$. Such an arrangement would require careful
matching of the beamsplitters $F$ and $G$ beforehand.

\begin{figure}[t]
\centerline{\includegraphics[width=3in]{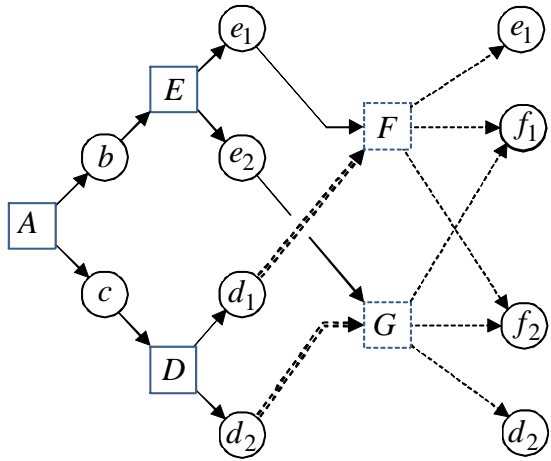}}
\caption{Proposed interfering beamsplitter experiment.}
\label{fig:6}
\end{figure}

The first two stages are as in Experiment 2, i.e., we have $\Psi
_{1}=(E_{1}e_{1}+E_{2}e_{2})(D_{1}d_{1}+D_{2}d_{2})$. We shall employ the
same method as in Experiment 4 to parametrize the transition from complete
no-which-path information to complete which-path information. We write%
\begin{eqnarray}
e_{1}d_{1} &\rightarrow &(\cos (\theta _{1})+\sin (\theta
_{1})u)(F_{1}f_{1}+F_{2}f_{2}),  \notag \\
e_{1}d_{2} &\rightarrow &e_{1}d_{2},  \notag \\
e_{2}d_{1} &\rightarrow &e_{2}d_{1},  \notag \\
e_{2}d_{2} &\rightarrow &(\cos (\theta _{2})+\sin (\theta
_{2})v)(F_{3}f_{1}+F_{4}f_{2}),
\end{eqnarray}%
where the $\{F_{i}\}$ satisfy the unitarity relations discussed earlier.
Hence we find%
\begin{eqnarray}
\Psi _{2} &=&(E_{1}D_{1}F_{1}\cos (\theta _{1})+E_{2}D_{2}F_{3}\cos (\theta
_{2}))f_{1}+  \notag \\
&&(E_{1}D_{1}F_{2}\cos (\theta _{1})+E_{2}D_{2}F_{4}\cos (\theta _{2}))f_{2}+
\notag \\
&&E_{1}D_{1}\sin (\theta _{1})u(F_{1}f_{1}+F_{2}f_{2})+E_{1}D_{2}e_{1}d_{2}
\notag \\
&&E_{2}D_{2}\sin (\theta _{2})v(F_{3}f_{1}+F_{4}f_{2})+E_{2}D_{1}e_{2}d_{1},
\notag \\
&&\
\end{eqnarray}%
which gives the total probabilities $P(f_{1}),P(f_{2})$ at ESDs $f_{1}$ and $%
f_{2}$ respectively to be%
\begin{eqnarray}
P(f_{1}) &=&|E_{1}D_{1}F_{1}|^{2}+|E_{2}D_{2}F_{3}|^{2}  \notag \\
&&+\left\{
\begin{array}{c}
E_{1}D_{1}F_{1}E_{2}^{\ast }D_{2}^{\ast }F_{3}^{\ast }\ \ \ \ \ \ \ \ \ \
\\
\ \ \ \ \ +E_{1}^{\ast }D_{1}^{\ast }F_{1}^{\ast }E_{2}D_{2}F_{3}%
\end{array}%
\right\} \cos (\theta _{1})\cos (\theta _{2}),  \notag \\
P(f_{2}) &=&|E_{1}D_{1}F_{2}|^{2}+|E_{2}D_{2}F_{4}|^{2}  \notag \\
&&+\left\{
\begin{array}{c}
E_{1}D_{1}F_{2}E_{2}^{\ast }D_{2}^{\ast }F_{4}^{\ast }\ \ \ \ \ \ \ \ \ \
\\
\ \ \ \ \ +E_{1}^{\ast }D_{1}^{\ast }F_{2}^{\ast }E_{2}D_{2}F_{4}%
\end{array}%
\right\} \cos (\theta _{1})\cos (\theta _{2}).  \notag \\
&&\
\end{eqnarray}

In the limit where there is complete which-path detection, i.e., $\theta
_{1}=\pi /2$ or $\theta _{2}=\pi /2$, we recover the results of Experiment $%
2 $, taking into account that $f_{1}$ and $f_{2}$ each then have to be
regarded as two ESDs. In the limit of complete no-which-path information,
i.e., $\theta _{1}=\theta _{2}=0$, we expect to observe interference effects
at $f_{1}$ and $f_{2}$.

The only real issue is whether all which-path information involving
beam-splitters $F$ and $G$ could be eliminated by mechanical means
sufficiently to produce interference. By this we mean removing all traces of
which beamsplitter had been triggered, equivalent to an $``$information
black hole$"$. This would undoubtedly require controlling the interaction
between the triggering beamsplitters $F$ and $G$ and their environment,
perhaps by intense cooling and shielding. The time scale for such erasure
would undoubtedly be a critical factor also.

It is possible that no technology could be devised to erase all traces of
beamsplitter switching, for both practical and theoretical reasons.
Theoretically, resetting a triggered beamsplitter to its untriggered state
amounts to resetting a pointer, and this is expected to carry a cost in
terms of irreversibility. The discussion in \cite{GREENBERGER+YASIN-1989} is
relevant here.

If it really were the case that all information concerning the triggering of
$F$ or $G$ could never be erased, then interference at $f_{1}$ and $f_{2}$
should never be observed, according to the rules of quantum mechanics. This
would mean that there really were two types of erasure. The first can be
called \emph{quantum erasure}, examples of which are the double-slit
experiment and the experiment discussed by Walborn et al. \cite{WALBORN-2002}%
. Interference can be observed in such cases. The second type of erasure,
\emph{classical erasure}, requires physical intervention on the part of the
observer. The big question then is whether classical and quantum erasure are
fundamentally different or not.

We believe that an experiment along the lines of our Experiment 5 could be
viable with current technology, but it would not be easy. A sequence of
steps would be taken, attempting to remove with more and more efficiency and
completeness any trace of which beamsplitter had been triggered,
corresponding to $\theta _{1}$ and $\theta _{2}$ both approaching $0$. It is
our intuition, based on the double-slit experiment and the experiment of
Greenberger and YaSin \cite{GREENBERGER+YASIN-1989}, that a point should
come where interference at $f_{1}$ and $f_{2}$ started to manifest itself,
but where that point is and whether it is attainable in practice are open
questions. Certainly, it would be an interesting experiment to attempt: even
the slightest hint of interference at $f_{1}$ and $f_{2}$ would cast light
on the fundamental questions $``$\emph{is there a Heisenberg cut?}$"$ and $%
`` $\emph{if there is such a cut, where does it start?}$"$.

\end{document}